\documentclass[aps,prl,reprint, amsmath, amssymb,superscriptaddress]{revtex4-1}

\usepackage[colorlinks=true,linkcolor=blue,urlcolor=blue,citecolor=blue,pdfusetitle]{hyperref}
\usepackage{bm}
\usepackage[retainorgcmds]{IEEEtrantools}
\usepackage{graphicx}
\usepackage{mathrsfs}
\usepackage{amsmath}
\usepackage{amssymb}
\usepackage{color}
\usepackage{amsfonts}
\usepackage{nicefrac}
\usepackage{nicefrac}
\usepackage{pifont}

\newcommand{\tr}{\mathrm{Tr}}

\begin{document}

\title{Extracting Bayesian networks from multiple copies of a quantum system}
\author{Kaonan Micadei}
\affiliation{Institute for Theoretical Physics I, University of Stuttgart, D-70550 Stuttgart, Germany}
\author{Gabriel T. Landi}
\affiliation{Instituto de F\'isica da Universidade de S\~ao Paulo,  05314-970 S\~ao Paulo, Brazil}
\author{Eric Lutz}
\affiliation{Institute for Theoretical Physics I, University of Stuttgart, D-70550 Stuttgart, Germany}

\begin{abstract}
Despite their theoretical importance,  dynamic Bayesian networks associated with quantum processes are currently not accessible experimentally. We  here describe a general scheme to determine the multi-time path probability of a Bayesian network  based on local measurements on independent copies of a composite quantum system combined with postselection. We further show that this protocol corresponds to a  non-projective measurement. It thus allows the investigation of the multi-time properties of a given local observable while fully preserving all its quantum features.
\end{abstract}

\maketitle 
Dynamic Bayesian networks offer a powerful framework to analyze conditional dependencies in a set of time-dependent random quantities. In this approach, relationships between dynamical variables are specified through conditional probabilities evaluated via \mbox{Bayes' rule~\cite{gha01,nea03,dar09,kos09}}. They have found widespread application in statistics, engineering and computer science  to model time series in probabilistic models. Hidden Markov models and Kalman filters are special cases of such \mbox{networks~\cite{gha01,nea03,dar09,kos09}}. In the past decade,  Bayesian networks have been successfully employed to investigate the nonequilibrium thermodynamics of small, composite systems, both in the classical \cite{sag12,ito13,kaw13,ehr17,cro19,str19,wol20,wol20a} and quantum \cite{mic20,par20,mic20a,str20} regimes. They have, in particular, been used to obtain fluctuation theorems,    fundamental generalizations of the second law  that characterize fluctuations of the entropy production arbitrarily far from equilibrium \cite{sei12}, for multipartite systems \cite{sag12,ito13,kaw13,ehr17,cro19,str19,wol20,wol20a,mic20,par20,mic20a,str20}. 

An interesting property of dynamic Bayesian networks is that they allow  to specify  the  local dynamics of a composite quantum system conditioned on its global state. The Bayesian network formalism thus preserves all the quantum features of the system, especially quantum correlations \cite{hor09} and quantum coherence \cite{str17}. As a result, it permits to go  beyond the standard two-projective-measurement scheme \cite{tal07,esp09,cam11}, which, owing to its projective nature, destroys off-diagonal density matrix elements. This characteristic has recently been exploited to  derive fully quantum fluctuation theorems that not only account for the quantum nonequilibrium dynamics of a driven system \cite{jar15}, as in  the  two-projective-measurement approach, but also fully capture both quantum correlations and quantum coherence at arbitrary times \cite{mic20,par20,mic20a}. However, while  a number of methods to implement the two-projective-measurement approach (and its variants) have been both  theoretically developed \cite{hub09,maz13,dor13,ron14} and experimentally  demonstrated \cite{bat14,an15,cer17,zha18,smi18,pal18,gom19}, to date,  no such protocol  exists for dynamic Bayesian networks.

In this paper, we introduce a general experimental scheme to extract dynamic Bayesian networks using   identical copies of a quantum system.
The use of multiple copies has been popularized   in quantum information theory to achieve, for example, entanglement detection \cite{hor03,min05,aol06,wal06,sch08,dal12,aba12,isl15,sag19} or  quantum state estimation \cite{eke02,bag05,bov05,pre06,ben09,llo14}, and has  recently been considered in quantum thermodynamics to reduce quantum backaction \cite{per17,wu19}. In the following, we first employ independent copies of a quantum system combined with postselection \cite{jac14} to reconstruct the path probability of a dynamic Bayesian network. The latter quantity determines the multi-time properties of a given local observable without requiring   full state tomography, which is in general extremely costly to realize \cite{ari03}. We moreover introduce  a positive-operator-valued measure (POVM)  \cite{jac14} such that the path probability directly results from global measurements of correlated copies in a broadcast state \cite{pia08}. We further show that a no-go  theorem for the characterization of work fluctuations in coherent quantum systems discussed in Ref.~\cite{per17} does not apply to such a POVM. A well-defined nonequilibrium quantum work distribution may consequently  be obtained for driven systems with initial coherence. We finally illustrate our findings by concretely evaluating the two-point path probabilities for a coherent qubit and for a quantum correlated pair of qubits.  

\emph{Dynamic Bayesian networks.} We consider an isolated quantum system initially prepared in a (global) state with spectral decomposition, $\rho = \sum_{s} P_s |s \rangle\langle s|$.
The system may be multipartite or single-partite. 
During its unitary evolution, $\rho_t =U_t \rho U_t^\dagger$, the  populations $P_s$ remain constant and the  basis elements rotate from $|s\rangle$ to $|s_t\rangle=U_t |s\rangle$. Let us now introduce arbitrary (local) basis sets $\{|x_0\rangle\}, \{|x_1\rangle\}, \ldots, \{|x_{N}\rangle\}$ at $(N+1)$ specific points in time, $t = t_0, t_1, \ldots, t_{N}$ (Fig.~1). These bases are not necessarily compatible with the bases $\{|s_{t} \rangle\}$ in the presence of quantum correlations or quantum coherence.

\begin{figure*}[t]
    \centering
    \includegraphics[width=0.8\textwidth]{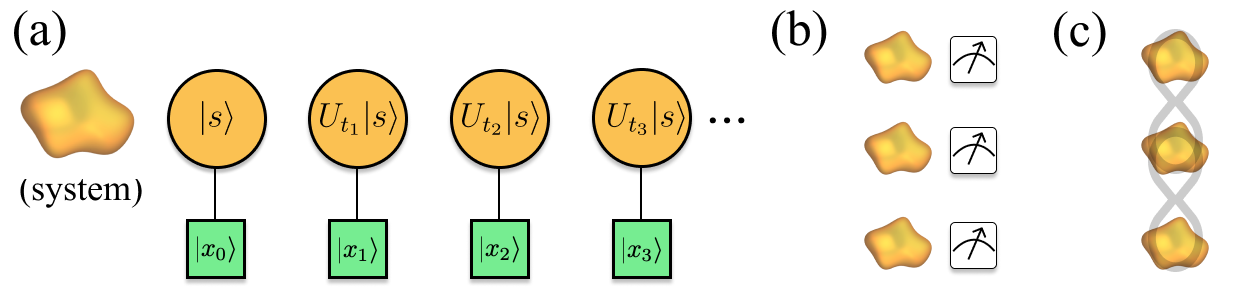}
    \caption{
    (a) Diagrammatic representation of the unitary evolution of a (possibly composite) quantum system  with (global) states $|s_t\rangle = U_t |s\rangle$, and the set of all possible (local) paths $|x_0\rangle \to |x_1\rangle \to |x_2\rangle$ which can be associated with this evolution.
    (b) The path probability $P(x_0,x_1,\cdots,x_N)$, Eq.~\eqref{prob_BN} may determined by performing local measurements $M_x$ on independent copies of a quantum system and postselecting the outcomes, Eq.~\eqref{post}.    
    (c) Alternatively, one may obtain the same statistics by performing a global measurement on  correlated copies  prepared in a broadcast state, Eq.~\eqref{broadcast}. 
    }
    \label{fig:diagram}
\end{figure*}

The central quantity of a dynamic Bayesian network is the joint probability distribution \cite{gha01,nea03,dar09,kos09}
\begin{equation}\label{prob_BN}
P(x_0,x_1,\cdots,x_N) = \sum\limits_{s} P_s \prod_{n=0}^{N}p(x_n|s_n),
\end{equation}
associated with a (local) path  $|x_0 \rangle \to |x_1\rangle \to |x_2\rangle \to \cdots$. The conditional probability of finding the system in the (local) state $|x_t\rangle$ given that it is in the (global) state  $|s_t\rangle$  at time $t$ is  $p(x_t|s_t) = |\langle x_t | U_t | s\rangle|^2$ \cite{mic20}. Equation \eqref{prob_BN} is a sum over all (global) trajectories $s$ of the path probability $P_s \prod_n p(x_n|s_n)$ of the conditional trajectory  $(s,x_0,x_1, \cdots, x_N)$. It is a proper probability distribution, in the sense that it is non-negative and all its marginals are non-negative. 
It also contains the complete information about the multi-time properties of the (local) variable $x$, while fully preserving the quantum features of the  system, in contrast to the two-projective measurement scheme \cite{tal07,esp09,cam11}.  It is the key quantity involved in the study of the nonequilibrium properties of small composite systems \cite{sag12,ito13,kaw13,ehr17,cro19,str19,wol20,wol20a,mic20,par20,mic20a,str20}. We next describe an experimental protocol to determine Eq.~\eqref{prob_BN} based on multiple identical copies of the quantum system and postselection.

  \emph{Experimental scheme.} We begin, for simplicity, by treating the case of the two-point  distribution $P(x_0,x_1)$ at time $t_0 =0$ and a later time $t_1$. To this end, we consider two independent copies $\rho \otimes \rho$ of the system.  We assume, as done  in the two-projective-measurement scheme, that the eigenbasis of the system has been determined.  The protocol  consists of two stages: In a first step, each copy is measured in the (global) eigenstate $|s\rangle$ by applying the   projector $\Pi_s \otimes \Pi_s $ with  $\Pi_s = |s\rangle\langle s|$. This results in the state $\left( \Pi_s \otimes \Pi_s \right) (\rho \otimes \rho) = P_s^2 \,\Pi_s \otimes \Pi_s$. In a second step, half of the copies are projected at $t= t_0$ in the (local) state $|x_0\rangle$, while the second half is projected  at $t= t_1$ in the (local) state $|x_1\rangle$, for a given (global) state $|s\rangle$. The corresponding  measurement operator
reads $M_{{x}} = |x_0\rangle\langle x_0 | \otimes U_{t_1}^\dagger |x_1\rangle\langle x_1| U_{t_1}$ and we obtain
\begin{equation}
	\label{eq:post-select}
	  \!\!\frac{\mathrm{Tr} \left[ M_x \left( \Pi_s \otimes \Pi_s \right) (\rho \otimes \rho) \right]}{\mathrm{Tr} \left[ \Pi_s \rho     \right]} 
        =P_s |\langle x_0 | s \rangle|^2 |\langle x_1 | U_{t_1} | s \rangle|^2
	\end{equation}
The joint probability distribution $P(x_0,x_1)$ of the dynamic Bayesian network then follows by summing Eq.~\eqref{eq:post-select} over all (global) trajectories $s$. 
We emphasize that  this protocol only relies on local measurements of each copy.
Moreover, since two-projective-measurement experiments already determine distributions by repeating measurements on many identically prepared systems \cite{bat14,an15,cer17,zha18,smi18,pal18,gom19}, the above  scheme may be realized without much additional  experimental effort.

The generalization  to an arbitrary sequence of times, $t_0, t_1, \cdots, t_N$, is straightforward. It involves $(N+1)$ independent copies, $\rho_\text{ind} =   \otimes_n\rho$, and the measurement operator $M_{{x}} = \otimes_n  (U_{t_n}^\dagger |x_n\rangle\langle x_n| U_{t_n})$. In this case, the multipoint joint probability distribution  \eqref{prob_BN} is  given by
\begin{equation}\label{post}
P(x_0,\cdots,x_N)=\sum_s \frac{1}{\tr(\Pi_s\rho)} \tr\Big[M_{{x}} (\otimes_n \Pi_s)  \rho_\text{ind} \Big]. 
\end{equation}
The  path probability \eqref{prob_BN} is thus obtained from  the  conditional expectation value of $M_{{x}}$ on postselected states.

   \emph{Generalized measurement operators.} The most general measurements in quantum theory are so-called positive-operator-valued measures (POVMs) \cite{jac14}. Such quantum measurements may always  be realized as ordinary projective measurements on an enlarged system  \cite{jac14}. In order to derive the POVM corresponding to the measurement of the path probability  \eqref{prob_BN}, we note that Eq.~\eqref{post} may be written as the expectation value
  \begin{equation}\label{broadcast}
    P(x_0,  \cdots,x_N) = \tr\Big[ M_{{x}}~\rho_\text{bro} \Big],
\end{equation} 
  where $\rho_{\text{bro}} = \sum_s P_s |s \cdots s\rangle\langle s \cdots s|$ denotes  a  broadcast state \cite{pia08}. Like the case of $(N+1)$ independent copies, this state  has the property  that if we take the partial trace over all except one of the subsystems, we always recover the original state $\rho$. Thus, locally, each copy is in state $\rho$, although, globally, they are in a quantum-correlated state.  The multipoint joint probability distribution  \eqref{prob_BN} of a dynamic Bayesian network may therefore be evaluated either using independent copies and post-selection or directly  as the outcomes  of the operator $M_x$ on a broadcast state of correlated copies.

We now introduce a completely positive trace-preserving map, $\mathcal{E}(\bullet) = \sum_i E_i \bullet E_i^\dagger$, with Kraus operators $E_{i} = \sum_{r} |r r \cdots r \rangle \langle r {i_1} \cdots {i_{N}}|$ and collective index $i= ( {i_1} \cdots i_{N})$ labeling the eigenstates of the system, such that the broadcast state can be constructed from $(N+1)$ independent copies as $\rho_\text{bro} = \mathcal{E}(\otimes_n\rho)$ \cite{com}. Using the cyclic property of the trace, we  obtain
\begin{equation}\label{broadcast_2}
    P(x_0,\cdots,x_N) = \tr\Big[J_{{x}} \big(\otimes_n\rho\big)\Big],
\end{equation}
with the  positive semidefinite operators $ J_{{x}} = \sum_{i} E_i^\dagger M_{{x}} E_i$. 
Since  $\sum_{{x}} J_{{x}} = 1$, they form a POVM \cite{jac14}. The set of operators $J_x$ define the general quantum measurement of the  path probability \eqref{prob_BN}  of a dynamic Bayesian network on $(N+1)$ independent copies.

 \emph{Connection with a no-go theorem for quantum work.}
Reference~\cite{per17} has recently examined general measurement schemes to evaluate the statistics of   nonequilibrium work performed on coherent systems. In this instance, the observable $x$ is the energy of the system and the work distribution is given as the expectation $P(w)= \tr[(\otimes_n \rho) W(w)]$, with the general work POVM $W(w) = \sum_{ij} \delta [w-(x_j -x_i)] J_x$.
The main conclusion of Ref.~\cite{per17} is that no POVM exists such that 
(i) the average work corresponds to  the difference of average energy  for closed quantum systems (first law) and (ii) the work statistics agree with the  two-projective-measurement method for states with no coherence in the energy basis (classical state limit), even if multiple copies are accessible. In other words,  it does not seem possible to simultaneously obey the first law of thermodynamics and respect the classical-state limit  in coherent systems. However, this result is based on the assumption that the measurement operator does not depend on the state $\rho$, that is, no information about the initial state is available. By contrast, we have here shown that  the  Bayesian-network approach allows the determination of the joint probability distribution \eqref{prob_BN}, and, in turn, of the nonequilibrium work distribution for  coherent (as well as  correlated multipartite) systems, by  relaxing this restriction and assuming that the eigenbasis of the system has been determined. In a sense, the hypothesis of state independence, which was based on a universality argument,  thus seems too strong. As a matter of fact, even the evaluation of the classical work statistics along single trajectories in stochastic thermodynamics presupposes knowledge of the driven potential \cite{sei12}. In addition, there exists many quantum protocols that require information about the eigenbasis of the system, from the two-projective-measurement scheme \cite{tal07,esp09,cam11} to optimal cloning \cite{bru98,sca05} and quantum parameter estimation \cite{bra94,mic15}.  

Compared with other  methods to specify  quantum work distributions, such as the two-projective-measurement scheme \cite{tal07,esp09,cam11}, the work-operator formalism \cite{all05,eng07} or  the quasiprobability approach \cite{all14,los18}, the dynamic-Bayesian-network framework \cite{mic20,par20,str20} appears to be  currently the only  one leading to quantum work distributions that (i) are measurable, that is, are described by a POVM, (ii) satisfy nonequilibrium fluctuation theorems and (iii) apply to coherent systems (Table I) \cite{per17}. It hence comes across as a powerful tool to study nonequilibrium quantum processes of composite systems.

\begin{table}[b]
\caption{Comparison of  different approaches to characterize  work fluctuations  in driven quantum systems: two projective measurements \cite{tal07,esp09,cam11}, work operators   \cite{all05,eng07},  quasiprobabilities \cite{all14,los18} and Bayesian networks \cite{mic20,par20,str20}. Only the latter one yields work densities that are measurable, obey fluctuation relations and apply to coherent systems.}
\begin{center}
\begin{ruledtabular}
\begin{tabular}{lccc}
& & Fluctuation & Coherent\\
&Measurable&  theorems&  processes\\
\hline
Projective measurements &\ding{51}& \ding{51}&\ding{55}\\
Operators of work &\ding{51}&\ding{55}&\ding{51}\\
Quasiprobabilities &\ding{55} & \ding{51} &\ding{51}\\
Bayesian networks &\ding{51} &\ding{51} & \ding{51}

\end{tabular}
\end{ruledtabular}
\end{center}
\end{table}

\emph{Examples.} We finally illustrate our results by computing the two-point path probability  \eqref{prob_BN} for two paradigmatic thermodynamic examples  for work extraction \cite{hek13} and heat exchange \cite{mic19}: a driven coherent qubit and  a correlated pair of qubits at  two different temperatures.

A) Driven coherent qubit. We consider the minimal example of a qubit with Hamiltonian $H_t = g_t \sigma_z$, whose gap  is adiabatically varied  from  $g_{t_0}$ to $g_{t_1}$. The system is assumed to be initially  in  { state $\rho = \rho_\text{th} + \alpha\sigma_x$}, with parameter $\alpha$ and thermal distribution  $\rho_\text{th} = \exp(-\beta g_0 \sigma_z) / Z$; here $Z = \mathrm{Tr} [\exp(-\beta g_0 \sigma_z)]$ denotes the partition function at inverse temperature $\beta=1/T$. The qubit exhibits coherences in the energy basis when $\alpha\neq0$. As a consequence, the eigenbasis $|s_\pm\rangle$ differs from the energy basis $|x_\pm \rangle$: we have $|s_+ \rangle = \cos({\theta}/{2})|x_+\rangle + \sin({\theta}/{2}) |x_-\rangle$ and $|s_- \rangle = -\sin({\theta}/{2}) |x_+\rangle + \cos({\theta}/{2}) |x_-\rangle$,  
where $\tan\theta = \alpha/b$,  
with $b = \tr[\sigma_z \rho_\text{th}]$.
The corresponding probabilities are $P_{s_\pm} = (1 \pm \sqrt{a^2 + b^2})/2$, where $a$ and $\alpha$ are related via $\alpha=a\sqrt{1-b^2}/2$, $|a|\leq 1$.

\begin{figure}[t]
    \centering
    \includegraphics[width=0.47\textwidth]{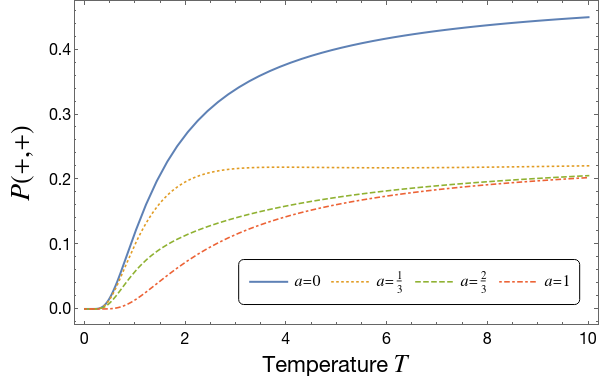}
    \caption{Path probability $P(+, +)$, Eq.~(6), for a driven coherent qubit as a function of  temperature $T$ for various values of the parameter $a$. The Bayesian network results $(a\neq 0)$ are generally smaller than that of the two-projective measurement scheme $(a=0)$, except for very low temperatures. They further plateau at a constant value for large temperatures.}
    \label{fig:work}
\end{figure}

During an adiabatic transformation only two paths occur: $|x_0 = +\rangle \rightarrow |x_1= +\rangle$ and  $|x_1 = -\rangle \rightarrow |x_1= -\rangle$. According to Eq.~\eqref{eq:post-select} the respective two-point path probabilities of the dynamic Bayesian network are
\begin{eqnarray}
  P(+, +)
        &=& p_{s_+}\,|\langle x_+ | s_+ \rangle|^2 |\langle x_+ |U_{t_1} | s_+ \rangle|^2\nonumber\\
        &+& p_{s_-}\,|\langle x_+ | s_- \rangle|^2 |\langle x_+ |U_{t_1} | s_- \rangle|^2\nonumber
        \\
      &=&\frac{1+ b}{2} -  \frac{{\alpha}^2}{4({\alpha}^2+b^2)},  \end{eqnarray}
and a similar expression for  $ P(-, -)$ with $b$ replaced with $-b$. Both formulas fully account for the presence of coherence ($a\neq0$) in  contrast to the two-projective-measurement approach, 
{in which the first measurement would completely destroy  coherences, effectively setting $a= 0$.
As a consequence, this would invariably lead to}
\begin{equation}
 P(+,+)_\text{TPM}
        = \langle x_+ | \rho | x_+ \rangle \,|\langle x_+ |U_{t_1} |x_+\rangle|^2
       = \frac{e^{-\beta g_0}}{Z},
\end{equation}
and, analogously, $ P(-,-)_\text{TPM} = {\exp({+\beta g_0})}/{Z}$. 
Figure 2 shows that quantum  coherence significantly affects the joint  distribution $P(+, +)$, except for {very low temperatures}: $P(+,+)$ is in general smaller than $P(+,+)_\text{TPM}$ and plateaus at a constant value at high temperatures.


B) Correlated pair of qubits. We next consider a pair of qubits $AB$ in the initial global state {$\rho_{AB} = \rho_\text{th}(\beta_A) \otimes \rho_\text{th}(\beta_B) + \alpha\sigma_+ \otimes \sigma_- + \alpha^*\sigma_- \otimes \sigma_+$} with $\rho_\text{th}(\beta_i) = \exp({-\beta_i \sigma_z}) / Z_i$ $(i=A,B)$, $\alpha = ia(Z_A Z_B)^{-1}$  and $|a|\leq 1$. The two qubits are initially correlated when $a\neq0$. As a consequence, the global eigenbasis $|s\rangle$  of  $\rho_{AB}$ differs from the local eigenbasis $|x\rangle=|\pm\pm\rangle$ of $\rho_A\otimes\rho_B$. 

The two qubits exchange energy during time $t_1$ by interacting  via a partial SWAP, $U_{t_1} = (I + iS)/\sqrt{2}$, where $S$ is the swap operator, $S|\phi \psi\rangle = |\psi \phi\rangle$. We thus  have $U_{t_1}|\pm \pm \rangle=  \exp({i\pi/4})|\pm \pm\rangle$ and 
 $U_{t_1}|\pm \mp\rangle =  ( |\pm \mp\rangle + i |\mp \pm\rangle)/\sqrt{2}$. We concretely compute the two-point joint probability distribution $P(+-, - +)$ of the dynamic Bayesian network for the local path $|+-\rangle \rightarrow |-+\rangle$ by evaluating the POVM 
   given in Eq.~\eqref{broadcast_2}. Using $M_{(+-)(-+)} = |+-\rangle\!\langle+-| \otimes U_{t_1}^\dagger |-+\rangle\!\langle-+| U_{t_1} $ and $E_i = \sum_r |rr \rangle\!\langle ri|$, with $|r\rangle$ and $|i\rangle$ eigenvectors of $\rho_{AB}$, we evaluate $J_{(+-)(-+)} = \sum_i E_i^\dagger M_{(+-)(-+)} E_i$ and obtain
   \begin{eqnarray}
       & J_{(+-)(-+)} 
	    = \frac{1}{2 \left\{4a^2 + \left[ \exp({-\Delta\beta}) - \exp({\Delta\beta}) \right]^2 \right\}} \times
	  \nonumber  \\
	    &\,\Big(
        |+ -\rangle\!\langle + -|\otimes \mathbf{A}
       + |+ -\rangle\!\langle - +|\otimes \mathbf{B}+
      \nonumber  \\
        &  |- +\rangle\!\langle + -|\otimes \mathbf{B}^\dagger
       + |- +\rangle\!\langle - +|\otimes \mathbf{C}
         \Big),
  \end{eqnarray}
with 
	$\mathbf{A} = \{ a^2 + \left[ \exp({-\Delta\beta}) - \exp({\Delta\beta}) - a \right]^2 \} I_4$,
$\mathbf{B} = -a \left\{ 2 i a+(1-i) \left[\exp({-\Delta \beta })-\exp({\Delta \beta})\right]\right\} I_4$ and 
$\mathbf{C} = 2a^2 \, I_4$. Taking the expectation value over two independent copies $ \rho_{AB}\otimes\rho_{AB}$, we eventually find
\begin{eqnarray}
\label{14}
        P(+-, - +) 
        &=& \mathrm{Tr} \left[ J_{(+ -)(- +)} \, \rho_{AB}\otimes\rho_{AB} \right] \nonumber
        \\
         = \frac{e^{-\Delta \beta }}{2 Z_A Z_B}&-&\frac{a \gamma }{Z_A Z_B \left[\gamma +e^{2 \Delta \beta } \left(e^{2 \Delta \
\beta
}+\xi \right)\right]} .
\end{eqnarray}
where we have defined $\xi = 2a^2 + 1$ and $\gamma = \exp({2\Delta\beta}) \xi + 1$.
Equation \eqref{14} entirely  captures   quantum correlations ($a\neq0$) between the  two qubits at $t_0$ and $t_1$, contrary  to the two-projective-measurement result to which it reduces for $a=0$.  Figure 3 displays the behavior of $P(+-, - +)$ as a function of $T_B$ for fixed $T_A$. We observe that quantum correlations have a nontrivial (nonmonotonic) influence on the path probability \eqref{prob_BN}. These effects vanish again in the limit of low temperatures.

\begin{figure}[t]
    \centering
    \includegraphics[width=0.47\textwidth]{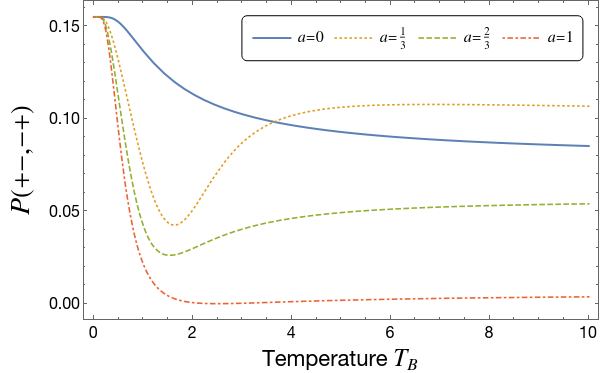}
    \caption{Path probability $P(+-, - +)$, Eq.~(9), for a pair of correlated qubits $AB$ as a function of  the temperature $T_B$ for various values of the parameter $a$ (and constant $T_A=0.4$). The Bayesian network results $(a\neq 0)$ strongly differ from  that of the two-projective measurement scheme $(a=0)$, except for very low temperatures, and exhibit nonmonotonic behavior.}
    \label{fig:heat}
\end{figure}

\emph{Conclusions.} We have introduced a general experimental scheme to extract a dynamic Bayesian network from multiple copies of  a multipartite quantum system. We have specifically shown how to determine the multipoint path probability \eqref{prob_BN}  from local measurements of independent copies combined with postselection. This joint probability density characterizes the multi-time properties of a given local observable, fully including quantum coherence and quantum correlations, without requiring  state tomography. We have further argued that this protocol may be regarded as a global generalized measurement and derived the corresponding POVM. In view of its versatility,  the present method can be implemented on many experimental platforms, including nuclear magnetic resonance \cite{bat14}, trapped ions \cite{an15}, cold atoms \cite{zha18} and superconducting qubits \cite{smi18}. We thus expect it to find broad applications in diverse fields, ranging from quantum many-body physics and quantum information theory to nonequilibrium quantum thermodynamics.

We acknowledge financial support from the S\~ao Paulo Research Foundation (Grants No. 2017/ 07973-5 and No. 2017/50304-7) and from the German Science Foundation (DFG) (Grant No. FOR 2724).


\begin{thebibliography}{99}
\bibitem{com} The quantum channel $\mathcal{E}$ requires knowledge of the eigenstates of $\rho$, which is why it does not violate the so-called no-broadcasting theorem~\cite{pia08}.
\end{thebibliography}
\end{document}